\documentclass[12pt]{article}
\usepackage{graphicx}
\usepackage{url}
\begin{document}

\def\b{\bigskip}
\def\s{\smallskip} \def\c{\centerline}  \baselineskip=15 pt  \c {\bf Is entanglement observer-dependent? }\b \c {Italo Vecchi} \c {Vicolo del Leoncorno 5 - 44100 Ferrara -
Italy} \c { email: vecchi@isthar.com } \b 

{ \bf Abstract}: {\sl The properties of quantum entanglement are examined and the role of the observer is pointed out.}\s
 

Entanglement plays a crucial role in current quantum theory, but  it is hard to find any clear
introduction to its properties. In this note some issues related to entanglement are examined and some conclusions are drawn, pointing out how  unphysical hidden assumption  have seeped  into the formalism used to describe entanglement. We refer to [1] for references and background material on entanglement, although some of our conclusions will differ from those proposed in the literature.  \s

The inadequacy of language in the description of quantum phenomena has been stressed by numerous authors.  The need to clarify the meaning of words is not new in scientific debate. Leibniz's letters to Clarke and Bohr's reply to the EPR paper [2] are classic texts where the search for the proper meaning of words plays a crucial role  . The dialogic form appears appropriate for such a purpose. Asking questions is an effective rethorical device and it may alert us to problems that may be hidden by a more reassuring formulation.\s

In the physical literature entanglement  appears for the first time in the classic EPR paper [2],
where the the entangled state $$
{ 1\over \sqrt 2}  ( |+\rangle_A  |-\rangle_B - |-\rangle_A  |+\rangle_B)
$$ of two spin-1/2 particles $ A $ and $ B $ is examined.

The most striking feature of entanglement is its non-local charachter: the measurement of the physical properties of one part of an entangled system appears to affect instantaneously other parts of the system. In the classic example of [2] the measurement of the spin of particle $ A $ appears to determine the value of the spin measurement of the entangled but spatially separated 
particle $ B $. In general we may say (Statement A) that "Two particles are entangled when the measurement of one of them affects the subsequent measurement of the other".  \s

A disturbing feature of Statement A is its ambiguity. If we identify a system with its state vector (i.e. with its wave-function), it is not clear "a priori"  how we can meaningfully refer to any of its components and what we mean by "particle".  We define particles through clusters of measurements, so that Statement A is intrinsecally circular. A good way around this problem is given by Statement B: "Two measurements are entangled if one of them affects the outcome of the other.".\s

At this point however there is still the problem of defining what constitutes a measurement. If we assume that the evolution of an isolated system is governed by a Schroedinger equation we must conclude that its evolution is  unitary. A salient feature of measurement is its irreversibility, while unitary evolution is always reversible. As long as the system can be imbedded in an isolated system its evolution is unitary and no irreversible change can take place. \s

There appears to be a contradiction here, but again we may find a way out considering the meaning of the words we are using. The key question in this setting is: "What constitutes an isolated system?". More precisely we may ask: "Isolated from what?". The easy answer is: "Isolated from the observer.".\s

Indeed we know that the act of observation induces state-vector reduction which implies loss of unitarity. This well-known fact underlies the popular "many-worlds" interpretation of quantum mechanics, first proposed by Everett. The problem is avoided only by the assumption of spontaneous state-vector reduction, i.e. of spontaneous loss of unitarity,  which we will not consider. \s

A relevant question at this point is: " When does state vector reduction takes place?". Mollifying answers based on dubious distinctions between the macroscopic and the microscopic level 
are quite popular, but such paths around the problem turn out to be circular. The straightforward 
answer to the above question is: "When the system under consideration ceases to be isolated from the observer".\s

Now a new question arises: "What consitutes an observer?". This is perhaps the hardest of all questions. A possible answer is "Human beings are observers". A frankly repulsive but operative one is "Physicists are observers".\s 

We can try to put things together. Based on the above we can claim that  "State-vector reduction takes place when the system ceases to be isolated from the observer".\s

We may reformulate the above as: "State-vector reduction takes place when the observer's perception of the system's state takes place". We may stop for a moment here. Perception is based on physical interaction. Without physical interaction there can be no perception. A system that is perceived interacts physically with the observer. If it interacts physically it is not isolated.
It somehow makes sense: "Unitarity is lost when the boundary between observer and observed breaks down, i.e. when perception takes place". Different perception may relie on different physical mechanisms, but without a physical interaction with the observer, e.g. a flow of photons hitting the observer's retina,  no breakdown of the the observer/observed boundary, i.e. no measurement,  can take place.\s

There is nothing really new here. We are just drawing the "Heisenberg cut" all the way back to the observer.\s

There is a thorny issue that must be tackled. State-vector reduction takes place in a certain basis in the system's state-space. Different bases yield different state-vector reductions. An obvious question is : "How is
the basis in which state-vector reduction takes place determined?". The answer is easy. A basis is a reference system in the system's state space. Reference systems are picked by the observer. In science there is no such thing as a physical system picking a basis, except perhaps in Tolemaic astronomy and certainly in decoherence theory ([3]). So the answer to the question is: "The observer picks the basis", i.e. "pointers" depend on the observer. Von Neumann may be nodding and adding: "The basis actually defines the observer". \s

Yes, but "why do different observers agree on their measurements?". In the Everett interpretation the question has been asked and answered in various ways, but mostly ignored . Maybe all observers we can communicate with are just instances of one observer, one mind, whatever. Maybe not all observers agree. Maybe certain groups of observers agree more among themselves than with other observers. Maybe there are clusters of observers with different degrees of relatedness. The less they are related, the harder it is for them to agree on what they observe and communicate effectively. We
may just accept the fact that, as far  as physical experiments are concerned, state-vector reduction relative to one experimenter appears to affect all experimenters. If it were not so, 
scientific communication, i.e. agreeing on observed facts, would be harder and perhaps impossible. \s

Let us sum up what we have got. Any physical system evolves as a wave-function governed by a Schroedinger equation and its evolution is unitary as long as no measurement is performed by the oberver. When the system is observed, its state-vector is projected into an observer-dependent basis. This projection process is in general  non-unitary. The observer-dependent nature of this process centers on the fact that the basis is observer-dependent. The amplitudes  of the system´s state-vector projections on the basis correspond to the probabilities of the possible outcomes. Probabilities measure the observer's ignorance of the measurement outcome in the observer's basis. In general we may note that probability, being a measure of ignorance, always refers to an observer.\s

We may have a look at the density matrix formalism in the light of the above remarks. Let us consider mixtures. In this setting the density matrix of mixtures represents the observer's knowledge on the possible outcomes of an experiment. It does not represent the state of the system.  Mixtures simply describe the observer´s knowledge of possible measurement outcomes in a certain basis.\s

We can now go back to entanglement and ask the question: "What is entanglement?" .The answer may be: "Entanglement is the observer's blueprint for state-vector reduction".
It should be clear that entanglement can be defined only in terms of the observer-dependent basis. Prior to observation all bases are equivalent so that speaking about entanglement is meaningless. It is only when state-vector reduction takes place that the system's state-vector is cast according to an observer-dependent set of rules . Entanglement has an observer-independent support, since the observer's perceptions are based on the information it extracts from its interaction with the system's state vector, which is determined by the system's evolution. However for state-vector reduction the physical features of the system, as encoded in the system's state-vector, must be interpreted  through a blueprint that depends on the observer. Loosely speaking we may say that physical interaction,
as described by the relevant Schroedinger equation,  may leave "marks" on the system`s state-vecto affecting the measurement outcome, e.g. the scrambling/vanishing of superpositions, but such "marks" are read according to an observer-dependent blueprint only when state-vector reduction takes place. Without an observer the "marks" are meaningless ripples on the system's wave-function.\s

It follows that when we use the standard notation relative to entanglement we are not describing the state of the system, but a set of possible measurement outcomes. This means that in the above EPR example the apparently harmless notation $ |+\rangle_A  |-\rangle_B $ does not refer to any intrinsic property of the system's state vector, but it indicates how state-vector reduction will be enforced on the system. The "product state" $ |+\rangle_A  |-\rangle_B $ in EPR is not a state at all, but just a notational convention on measurement outcomes, expressing how subsequent measurements will be related. Actually the requirement that total spin is conserved is a constraint on measurement outcomes, i.e. on spin measurements. Such a constraint is meaningless as a condition on the system's state-vector.  
This distinction is not purely formal, since prior to observation the system may evolve in a way that may not be consistent with the formalism  describing entanglement. Such a discrepancy may be experimentally detected ([4]). 

\b

 \c{ \bf References} \b

[1] P.Kwiat, H.Weinfurter, T.Herzog and A. Zeilinger, Phys. Rev. Lett. 74(24), 4763 (1995) \s

[2] A.Einsten, B.Podolsky and N.Rosen, Phys. Rev. 47, 777 (1935) \s

[3] I.Vecchi, at \url{http://xxx.lanl.gov/abs/quant-ph/0002084} (see also: \url{http://xxx.lanl.gov/abs/quant-ph/0102130}). \s
 
[4] I.Vecchi, at \url{http://xxx.lanl.gov/abs/quant-ph/0007117} \s
 \s

\end{document}